\documentclass[aps,prd,onecolumn,groupedaddress,showpacs,showkeys,nofootinbib,amssymb]{revtex4}
\usepackage[latin1]{inputenc}
\usepackage{graphicx}
\usepackage[english]{babel}
\usepackage{amsmath}
\usepackage{amssymb}
\usepackage{amsfonts}

\begin{document}
\def\pp{{\, \mid \hskip -1.5mm =}}
\def\cL{{\cal L}}

\def\be{\begin{equation}}
\def\ee{\end{equation}}
\def\bea{\begin{eqnarray}}
\def\eea{\end{eqnarray}}
\def\beq{\begin{eqnarray}}
\def\eeq{\end{eqnarray}}
\def\tr{{\rm tr}\, }
\def\nn{\nonumber \\}
\def\e{{\rm e}}

\title{\textbf{Testing metric-affine $f(R)$-gravity   by  relic scalar gravitational waves}}

\author{S. Capozziello$^{1,2}$, R. Cianci$^{3}$, M. De Laurentis$^{1,2}$, S. Vignolo$^{3}$}

\affiliation{$~^{1}$ Dipartimento di Scienze Fisiche,
Universit\`{a} ``Federico II'' di Napoli, Compl. Univ. Monte S.
Angelo Ed. N, via Cinthia, I- 80126 Napoli (Italy),}
 \affiliation{$^{2}$INFN Sez. di
Napoli, Compl. Univ. Monte S. Angelo Ed. N, via Cinthia, I- 80126
Napoli (Italy),} \affiliation{$^{3}$DIPTEM Sez. Metodi e Modelli
Matematici, Universit\`a di Genova,  Piazzale Kennedy, Pad. D -
16129 Genova (Italy).}

\date{\today}

\begin{abstract}
We discuss the emergence of scalar  gravitational waves in  metric-affine
$f(R)$-gravity.  Such a component allows to discriminate between metric and metric--affine theories  %{\it \`a
%la\/} Palatini  and with torsion.
The intrinsic meaning of this result is that the geodesic structure of the theory can be discriminated.
%In fact the presence of torsion  means that the connection is not Levi-Civita but it is more general.
We extend the formalism of  cross
correlation analysis, including the additional polarization mode,
and calculate the detectable energy density of the spectrum for
cosmological  relic gravitons.  The possible detection of the signal is discussed  against  sensitivities of  VIRGO, LIGO and LISA interferometers.
\end{abstract}

\maketitle
\vspace{5.mm}
%%%%%%%%%%%%%%%%%%%%%%%%%%%%%%
\section{Introduction}
\label{intro}
%%%%%%%%%%%%%%%%%%%%%%%%%%%%%%%

Extending General Relativity (GR) to more general actions with respect to the
Hilbert-Einstein one is revealing a very fruitful approach in modern physics. From a
conceptual point of view, there is no {\it a priori} reason to
restrict the gravitational Lagrangian to a linear function of the
Ricci scalar $R$, minimally coupled with matter. The idea that
there are no ``exact'' laws of physics but that the Lagrangians of
physical interactions are ``stochastic'' functions -- with the
property that local gauge invariances (i.e. conservation laws) are
well approximated in the low energy limit and that physical
constants can vary -- has been recently taken into serious consideration.

Beside fundamental physics motivations, all these theories have
acquired a huge interest in cosmology due to the fact that they
``naturally" exhibit inflationary behaviors able to overcome the
shortcomings of Cosmological Standard Model.
Furthermore, dark energy models mainly rely on the implicit
assumption that Einstein's GR is the correct theory of gravity,
indeed. Nevertheless, its validity at the larger astrophysical and
cosmological scales has never been tested and it is
therefore conceivable that both cosmic speed up and missing matter
represent signals of a breakdown in our understanding of
gravitation law so that one should consider the possibility that
the Hilbert\,-\,Einstein Lagrangian, linear in the Ricci scalar
$R$, should be generalized. Following this line of thinking, the
choice of a generic function $f(R)$ can be derived by matching the
data and by the "economic" requirement that no exotic ingredients
have to be added. This is the underlying philosophy of what is
referred to as $f(R)$--gravity  \cite{reviews}.
 In this context, the same cosmological constant could be
removed as an ingredient of the cosmic pie being nothing else but
a particular eigenvalue of a general class of theories
\cite{garattini}.

However $f(R)$--gravity can be encompassed in the Extended Theories
of Gravity being a "minimal" extension of GR where
functions of Ricci scalar are taken into account. Although such gravity theories
have received much attention in cosmology,
since they are naturally able to give rise to  accelerating
expansions (both in the late and in the early Universe) and
systematic studies of the phase space of solutions are in progress
\cite{cnot}, it is possible to demonstrate that $f(R)$--theories
can also play a major role at astrophysical scales \cite{mnras}.

Despite these encouraging results, a major conceptual problem has
not been solved  indeed: the description of  dynamics can be
pursued in metric, affine and metric-affine approaches with
different results (see \cite{reviews} for details).  This means
that the affine connection cannot be simply Levi-Civita, implying
that the geodesic and causal structure coincide as in the metric
approach, but it could be endowed with a richer geometric
structure. For example, if the spacetime is ${\bf U}_4$ and not
${\bf V}_4$ as in Riemannian geometry, torsion fields have to be
considered and connection is not simply Levi-Civita \cite{CCSV1}.
The consequence of this fact is that geodesic structure and causal
structure of spacetime can be disentangled  \cite{hehl}. Probing
this feature is a fundamental issue that could have dramatic
consequences  since further degrees of freedom, like spin or the
different types of torsion, should be considered into dynamics
\cite{CLS}. Searching for experimental probes in this direction is
a urgent problem faced by several authors (see e.g.
\cite{guthtor}).

In this paper, we will take into account the metric-affine
formulation of  $f(R)$-gravity and put in evidence that the
difference between a theory  {\it \`a la\/} Palatini  and a theory
with torsion can be reconducted to the difference between affine
connections that results in the relative variation of a scalar
field. Such a scalar field is endowed with the further degrees of
freedom by which $f(R)$-gravity differs from GR where $f(R)=R$.
Assuming that the effect of such a field is a further scalar mode
in the relic cosmic background of gravitational waves (GWs),  we
consider that the possible detection of such a mode could result,
for example, as a probe of torsion. On the other hand, the absence
of such a component could be a confirmation that the connection is
actually Levi-Civita.  Besides,  detecting new gravitational modes
could be a sort of {\it experimentum crucis} in order to
discriminate among theories since this fact would be the
``signature" that GR should be enlarged or modified
\cite{maggiore,bellucci,elizalde}. The outline of the paper is as
follows.
 In Sec. \ref{tre}, we  give an overview of
metric--affine $f(R)$--gravity theories. %The main result of this section is that the presence of the torsion field can %be achieved by confronting torsion connection with Palatini connection.
The linearized theory of gravitational waves is considered in Sec.
\ref{quattro} . Here,  the standard polarization, modes coming
from GR, are distinguished with respect to the  further possible
scalar mode coming from the preceding considerations. In Sect.
\ref{sei}, we investigate the response of  a single detector to a
GW propagating in  certain direction with each polarization mode.
Sect. \ref{sette} is devoted to the  discussion of  the spectrum
of the  GW stochastic background where  the further scalar mode is
considered. Conclusions are drawn in Sect. \ref{otto}.

%%%%%%%%%%%%%%%%%%%%%%%%%%%%%%%%%%%%%%%%%%%%%
\section{Metric-affine $f(R)$-gravity}
\label{tre}
%%%%%%%%%%%%
In metric--affine $f(R)$- gravity, the
(gravitational)  dynamical fields are pairs $(g,\Gamma)\/$
consisting of a pseudo--Riemannian metric $g\/$ and a linear
connection $\Gamma\/$ on the space-time manifold $M\/$.
In the Palatini approach, the connection
$\Gamma\/$ is torsionless but it is not requested to be
metric--compatible, instead, in the approach with torsion, the
dynamical connection $\Gamma$ is forced to be metric but it is allowed to have
torsion different from zero.
The field equations are derived from an action functional of the form
\begin{equation}\label{2.1}
{\cal A}\/(g,\Gamma)=\int{\left(\sqrt{|g|}f\/(R) + {\cal L}_m\right)\,ds}
\end{equation}
where $f(R)$ is a real function, $R\/(g,\Gamma) = g^{ij}R_{ij}\/$
(with $R_{ij}:= R^h_{\;\;ihj}\/$)  is the scalar curvature
associated with the connection $\Gamma\/$ and ${\cal L}_m\,ds\/$
is a suitable material Lagrangian.
Assuming that the material Lagrangian does not depend on the dynamical connection, the field equations are
\begin{equation}\label{2.2a}
f'(R)R_{ij} - \frac{1}{2}f\/(R)g_{ij}=\Sigma_{ij}\,,
\end{equation}
\begin{equation}\label{2.2b}
T_{ij}^{\phantom{ij}h} = -
\frac{1}{2f'(R)}\frac{\partial{f'(R)}}{\partial{x^p}}\left(\delta^p_i\delta^h_j
- \delta^p_j\delta^h_i\right)
\end{equation}
for $f(R)$-gravity with torsion $T_{ij}^{\phantom{ij}h}\equiv
\Gamma^{\phantom{ij}h}_{ij}-\Gamma^{\phantom{ji}h}_{ji}$
\cite{CCSV1,CCSV2,CCSV3}, and
\begin{equation}\label{2.3a}
f'\/(R)R_{ij} - \frac{1}{2}f(R)g_{ij}=\Sigma_{ij}\,,
\end{equation}
\begin{equation}\label{2.3b}
\nabla_k(f'(R)g_{ij})=0\,,
\end{equation}

for $f(R)$-gravity in the  Palatini approach \cite{francaviglia1,Sotiriou,Sotiriou-Liberati1,Olmo}. In Eqs.
\eqref{2.2a} and \eqref{2.3a}, the quantity ${\displaystyle
\Sigma_{ij}:= -
\frac{1}{\sqrt{|g|}}\frac{\delta{\cal L}_m}{\delta g^{ij}}\/}$ plays the role
of stress-energy tensor. Considering the trace of Eqs.
\eqref{2.2a} and \eqref{2.3a}, we obtain the relation
\begin{equation}\label{2.4}
f'\/(R)R -2f\/(R) = \Sigma
\end{equation}
linking the curvature scalar $R\/$ with the trace of the stress-energy tensor
$\Sigma:=g^{ij}\Sigma_{ij}\/$.

From now on, we shall suppose that the relation \eqref{2.4}  is
invertible as well as that $\Sigma\neq const$ (this implies,
for example, $f(R)\/$ different from $\alpha R^2\/$ which is only
compatible with $\Sigma=0\/$) . Under these hypotheses, the
curvature scalar $R\/$ can be expressed as a suitable function of
$\Sigma\/$, namely
\begin{equation}\label{2.5}
R=F(\Sigma)\,.
\end{equation}
If $\Sigma=const$, GR plus the cosmological constant is recovered \cite{CCSV1}.
Defining the scalar field
\begin{equation}\label{2.6}
\varphi:=f'(F(\Sigma))
\end{equation}
we can put the Einstein--like field equations of both {\it \`a
la\/} Palatini  and with torsion theories in the same form
\cite{CCSV1,Olmo}, that is

\begin{eqnarray}\label{2.7}
 \tilde{R}_{ij} -\frac{1}{2}\tilde{R}g_{ij}=
 \frac{1}{\varphi}\Sigma_{ij}
+ \frac{1}{\varphi^2}\left( \varphi\tilde{\nabla}_{j}\frac{\partial\varphi}{\partial{x^i}} - \frac{3}{2}\frac{\partial\varphi}{\partial{x^i}}\frac{\partial\varphi}{\partial{x^j}}+ \frac{3}{4}\frac{\partial\varphi}{\partial{x^h}}\frac{\partial\varphi}{\partial{x^k}}g^{hk}g_{ij}
 - \varphi\tilde{\nabla}^h\frac{\partial\varphi}{\partial{x^h}}g_{ij} -
V\/(\varphi)g_{ij} \right)\,,\nonumber\\
\end{eqnarray}
where we have introduced the effective potential
\bigskip\noindent
\begin{equation}\label{2.8}
V\/(\varphi):= \frac{1}{4}\left[ \varphi
F^{-1}\/((f')^{-1}\/(\varphi)) +
\varphi^2\/(f')^{-1}\/(\varphi)\right]\,,
\end{equation}
for the scalar field $\varphi\/$. In Eq. \eqref{2.7}, $\tilde{R}_{ij}\/$, $\tilde{R}\/$ and $\tilde\nabla\/$ denote respectively the Ricci tensor, the scalar curvature and the covariant derivative associated with the Levi--Civita connection of the dynamical metric $g_{ij}$.
Therefore, if the dynamical connection $\Gamma\/$ is not coupled
with matter, both the theories (with torsion and Palatini--like)
generate identical Einstein--like field equations.

On the contrary, the field
equations for  dynamical connection are different and (in
general) give rise to different solutions. In fact, the connection
$\Gamma$ solution of Eqs. \eqref{2.2b} is
\begin{equation}\label{2.9}
\Gamma_{ij}^{\phantom{ij}h} =\tilde{\Gamma}_{ij}^{\phantom{ij}h} +
\frac{1}{2\varphi}\frac{\partial\varphi}{\partial{x^j}}\delta^h_i -
\frac{1}{2\varphi}\frac{\partial\varphi}{\partial{x^p}}g^{ph}g_{ij}
\end{equation}
where $\tilde{\Gamma}_{ij}^{\phantom{ij}h}\/$ denote the coefficients of
the Levi--Civita connection associated with the metric $g_{ij}$,
while the connection $\bar\Gamma\/$ solution of Eqs. \eqref{2.3b}
is
\begin{equation}\label{2.10}
\bar{\Gamma}_{ij}^{\phantom{ij}h}= \tilde{\Gamma}_{ij}^{\phantom{ij}h} +
\frac{1}{2\varphi}\frac{\partial\varphi}{\partial{x^j}}\delta^h_i -
\frac{1}{2\varphi}\frac{\partial\varphi}{\partial{x^p}}g^{ph}g_{ij} +
\frac{1}{2\varphi}\frac{\partial\varphi}{\partial{x^i}}\delta^h_j\,,
\end{equation}
and coincides with the Levi--Civita connection induced by the
conformal metric $\bar{g}_{ij}:=\varphi g_{ij}$. By comparison,
the connections $\Gamma$ and $\bar\Gamma$ satisfy the relation
\begin{equation}\label{2.11}
\bar{\Gamma}_{ij}^{\phantom{ij}h} -\Gamma_{ij}^{\phantom{ij}h} =
\frac{1}{2\varphi}\frac{\partial\varphi}{\partial{x^i}}\delta^h_j\,.
\end{equation}
Of course, the Einstein--like equations \eqref{2.7} are coupled
with the matter field equations. In this respect, it is worth
pointing out that Eqs. \eqref{2.7} imply the same conservation
laws holding in GR  \cite{CV1,CV2,CV3}. An important consideration is
in order at this point. The difference between the affine
connections in  theories with torsion and  {\it \`a la\/} Palatini
is characterized by the relative variation of the scalar field
$\varphi$, which is determined by the matter-energy $\Sigma$.
 Assuming such a difference as a perturbation in a fixed conformal background,
 we can investigate if a scalar mode of gravitational radiation could be related with it.
 This will be the argument of the next section.

%%%%%%%%%%%%%%%%
\section{The scalar mode of gravitational waves and polarization states}
\label{quattro}
%%%%%%%%%%%%%%%%%
%
The geodesic structure of metric-affine $f(R)$-gravity could be
tested by detecting a further polarization in gravitational
radiation related to a scalar mode.  To be more precise,  the
scalar field is a way to define the further gravitational degrees
of freedom emerging from $f(R)$-gravity. The total GW has to be
a function of the standard modes coming from GR and modes coming
from these further degrees of freedom, that is
$h^{(tot)}_{ij}=h^{(tot)}_{ij}(h_{ij}, h_{\varphi})$. The
following derivation will show this statement.

Let us consider a small perturbation on a conformal background
characterized by a Minkowskian spacetime  on which a scalar field
$\varphi=\varphi_{0}$ is defined.   Perturbing the background at
first order, we obtain

\begin{equation}
\begin{array}{c}
\tilde{g}_{ij}=\eta_{ij}+h_{ij}\,,\qquad\varphi=\varphi_{0}+\delta\varphi.\end{array}\label{eq:
linearizza}\end{equation}

From these linearized quantities, it is possible to derive the
related curvature invariants $\widetilde{R}_{ijkl}$,
$\widetilde{R}_{ij}$ and $\widetilde{R}$ and the field equations:

\begin{equation}
\begin{array}{c}
\widetilde{R}_{ij}-\frac{\widetilde{R}}{2}\eta_{ij}=(\partial_{i}\partial_{j}h_{\varphi}-\eta_{ij}\square h_{\varphi})\,,\qquad
{}\square h_{\varphi}=m^{2}h_{\varphi},\end{array}\label{eq:linearizzate1}\end{equation}
where
$\displaystyle{h_{\varphi}\equiv\frac{\delta\varphi}{\varphi_{0}}}$
(see \cite{CST} for details). On the other hand, such equations
can be obtained directly starting from the field Eqs.(\ref{2.7}) with the
potential (\ref{2.8}). It is straightforward to see that
$\displaystyle{\delta\varphi=\frac{\partial \varphi}{\partial
x^i}\delta x^i}$  being $\delta x^i=dx^i$. Looking at Eq.
\eqref{2.11}, it is clear that the relative variation of the
scalar field with respect to the background gives the difference
between the connections $\bar{\Gamma}$ and $\Gamma$ and then it is
the signature of the fact that connection is not Levi-Civita.
Besides, being the Klein-Gordon equation for the scalar field
$\varphi$ in Minkowski spacetime

\begin{equation}
\square\varphi=\frac{dV}{d\varphi}\,,\label{eq: KG2}\end{equation}
and perturbing Eq.\eqref{2.8}, we have at the lowest order

\begin{equation}
V\simeq\frac{1}{2}m^2\,\delta\varphi^{2}\Rightarrow\frac{dV}{d\varphi}\simeq
m^{2}\delta\varphi,\label{eq: minimo}\end{equation} and then the
above result in  \eqref{eq:linearizzate1}. The constant $m$ has
the dimensions of a mass. $\widetilde{R}_{ijkl}$ and Eqs.
(\ref{eq:linearizzate1}) are invariants for gauge transformations
\cite{maggiore}

\begin{equation}
\begin{array}{c}
h_{ij}\rightarrow h'_{ij}=h_{ij}-\partial_{(i}\epsilon_{j)}\,
,\qquad
\delta\varphi\rightarrow\delta\varphi'=\delta\varphi;\end{array}\label{eq:
gauge}\end{equation} then

\begin{equation}
\bar{h}_{ij}\equiv h_{ij}-\frac{h}{2}\eta_{ij}\label{eq:
ridefiniz}\end{equation} can be defined. Here $h$ is the trace. By
considering  suitable transformation parameters $\epsilon^{j}$,
one  gets an equivalent Lorentz gauge where the following
equations hold

\begin{equation}
\partial^{i}\bar{h}_{ij}=0\,.\label{eq: condlorentz}\end{equation}
According to this choice, field equations read

\begin{equation}
\square\bar{h}_{ij}=0\,, \qquad \square
h_{\varphi}=m^{2}h_{\varphi}\, .\label{eq: ondaS}\end{equation}
Solutions of Eqs.(\ref{eq: ondaS}) are plane waves:

\begin{equation}
\bar{h}_{ij}=A_{ij}(\overrightarrow{k})\exp(ik^{j}x_{j})+c.c.\,,\label{eq: solT}
\end{equation}

\begin{equation}
h_{\varphi}=a(\overrightarrow{k})\exp(iq^{j}x_{j})+c.c.\,,\label{eq:
solS}\end{equation} where the following dispersion relations hold

\begin{equation}
\begin{array}{ccc}
k^{j}\equiv(\omega,\overrightarrow{k})\, , &  & \omega=k\equiv|\overrightarrow{k}|\, ,\\
\\q^{j}\equiv(\omega_{m},\overrightarrow{k})\, , &  & \omega_{m}=\sqrt{m^{2}+k^{2}}.\end{array}\label{eq: keq}\end{equation}
The polarization tensor $A_{ij} (\overrightarrow{k})$ can be
found following the arguments in  Ref. \cite{vanDam:1970vg,greci}.
Here Eq. (\ref{eq: solT}) represents  the standard waves of GR
\cite{Misner,Allen}.
 Eq.  (\ref{eq: solS}) is  the solution for the massive mode.
The fact that the dispersion law for the modes of the massive
field $h_{\varphi}$ is not linear has to be stressed. The velocity
of every {}``ordinary'' mode $\bar{h}_{ij}$, arising from GR, is the light speed $c$, but the dispersion law (the
second of Eq. (\ref{eq: keq})) for the modes of $h_{\varphi}$ is
that of a massive field. It  can be discussed as a wave-packet propagating on the background.
Also, the group-velocity of a wave-packet of $h_{\varphi}$ centered in
$\overrightarrow{k}$ is

\begin{equation}
\overrightarrow{v_{G}}=\frac{\overrightarrow{k}}{\omega_m},\label{eq:
velocita' di gruppo}\end{equation}
which is exactly the velocity of a massive particle with mass $m$,
momentum $\overrightarrow{k}$ and frequency $f=\omega_{m}/2\pi$. From the second of Eqs.
(\ref{eq: keq}) and Eq. (\ref{eq: velocita' di gruppo}), it is
straightforward to obtain:

\begin{equation}
v_{G}=\frac{\sqrt{\omega_m^{2}-m^{2}}}{\omega_m}.\label{eq: velocita'
di gruppo 2}\end{equation} Then, assuming a constant speed for the
wave-packet, it has to be

\begin{equation}
m=\sqrt{(1-v_{G}^{2})}\omega_{m}.\label{eq: relazione massa-frequenza}\end{equation}
Now, it has to be discussed if there could be
phenomenological limitations to the GW-mass. A strong
limitation arises from the fact that the GW should be in  a frequency
which falls in the frequency-range for both Earth-based and
space-based gravitational antennas, that is the interval
$10^{-4}Hz\leq f\leq10KHz$
\cite{acernese,willke,sigg,abbott,ando,tatsumi,lisa1,lisa2}. For a
massive GW,  it is:

\begin{equation}
2\pi f=\omega_{m}=\sqrt{m^{2}+k^{2}}\,.\label{eq:
frequenza-massa}\end{equation}
Thus, it has to be

\begin{equation}
0\leq m\leq10^{-11}eV.\label{eq: range di massa}\end{equation}
A stronger limitation is given by the requirements coming from cosmology and
Solar System tests on modified theories of gravity (see e.g. \cite{maggiore}). In this case,
it is
\begin{equation}
0\leq m\leq10^{-33}eV.\label{eq: range di massa 2}\end{equation}
These scalar GW modes can be dealt as very light  particles.

Let us consider now the GW- polarization states. The above
gauge gives  $k^{i}A_{ij}=0$. Moreover, we assume that
 $\overrightarrow{k}$ is in the $z$ direction and we choose  a gauge in which only
$A_{11}=-A_{22}$, and $A_{12}=A_{21}$ are different from zero. In
this frame, we take a polarization base  of  the form

\begin {equation}
e_{ij}^{(+)}=\frac{1}{\sqrt{2}}\left(
\begin{array}{cccc}
0&0 & 0 & 0 \\
0&1 & 0 & 0 \\
0 & 0 & -1&0 \\
0 & 0 & 0&0
\end{array}
\right),\qquad e_{ij}^{(\times)}=\frac{1}{\sqrt{2}}\left(
\begin{array}{cccc}
0&0 & 0 & 0 \\
0&0 & 1 & 0 \\
0&1 & 0 & 0 \\
0&0 & 0 & 0
\end{array}
\right)\, ,\qquad e_{ij}^{(s)}=\frac{1}{\sqrt{2}}\left(
\begin{array}{cccc}
0&0 & 0 & 0 \\
0&0 & 0 & 0 \\
0&0 & 0 & 0 \\
0 & 0 & 0&1
\end{array}\right).
\end{equation}
The resulting GW is

\begin{equation}
h^{(tot)}_{ij}(t,z)=
A^{+}(t-z)e_{ij}^{(+)}+A^{\times}(t-z)e_{ij}^{(\times)}+h_{\varphi}(t-v_{G}z)e_{ij}^{s}.\label{eq:
perturbazionetotale}\end{equation} The first two terms
$A^{+}(t-z)e_{ij}^{(+)}$ and $A^{\times}(t-z)e_{ij}^{(\times)}$
describe the two standard polarizations of gravitational waves
that arise from GR, while the term
$h_{\varphi}(t-v_{G}z)e_{ij}^{s}$ is the massive field arising
from  $f(R)$-gravity. According to the derivation of previous
section, it could characterize the fact that connection is not
Levi-Civita and then the geodesic structure of the theory could be
distinguishable with respect to the causal structure. In Fig.1, we
illustrate how each GW polarization affects test masses arranged
on a circle.
\begin{figure}
\begin{center}
\leavevmode
\includegraphics[scale=0.6]{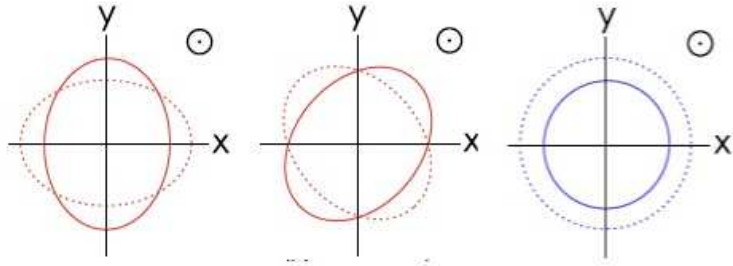}
\caption{The polarization modes of gravitational waves. In the
figure,  displacements  that each mode induces on a sphere of test
particles are shown.  We have respectively the standard plus and
cross modes of GR, and the scalar mode related to the scalar
field.  The circled dots represent the inward propagating
direction, orthogonal to the plane.}
\end{center}
\label{fig1}
\end{figure}
%%%%%%%%%%%%%%%%%%%%%%%%%%%%%%%%%%%%%%%%%%%
\section{The response of  interferometers to scalar gravitational waves}
\label{sei}
\begin{figure}[h]
\begin{center}
\includegraphics[width=6.5cm]{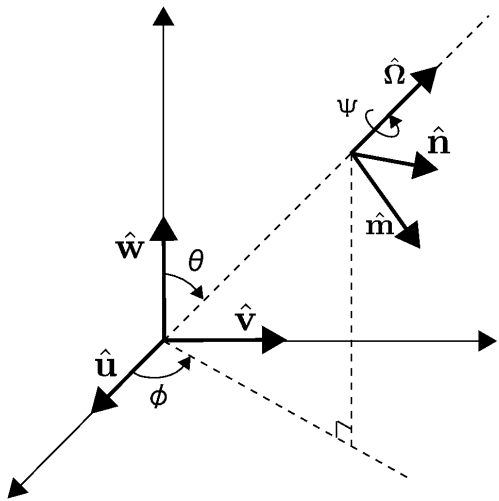}
\caption{The  coordinate systems used to calculate the
polarization tensors and the pictorial view of the coordinate
transformation.} \label{fig2}
\end{center}
\end{figure}

\begin{figure}[t]
\centering
{\vspace{0cm}\includegraphics[width=0.7\textwidth]{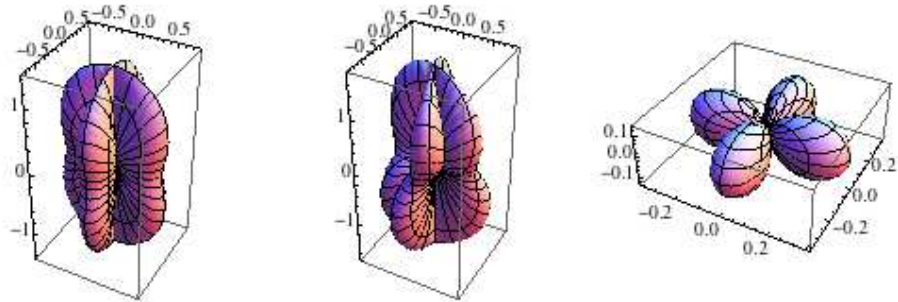}}
\caption{The angular pattern (i.e. the sensitivity) of interferometer for $+$, $\times$ and scalar polarizations. }
\label{fig:3}
\end{figure}

\begin{figure}[t]
\centering
{\vspace{0cm}\includegraphics[width=0.7\textwidth]{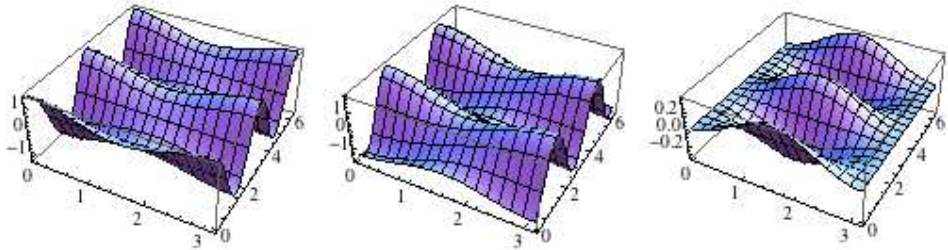}}
\caption{The propagation effects of  $+$, $\times$ and scalar
polarizations. } \label{fig:3a}
\end{figure}

%%%%%%%%%%%%%%%%%%%%%%%%%%%%%%%%%%%%%%%%%%%
Let us compute now the  response measured in an
interferometer when a GW is coming from an arbitrary direction.
We suppose that a further polarization, related to the scalar mode, is present so the antenna pattern should say, in principle, if the detection of such a mode is possible.
First of all, let us  construct  a  {\it response tensor or
detector tensor} $\mathbf{D}$ such that the signal induced in the
detector by a GW of polarization $ \mathbf{e}$ is proportional to
the angular pattern function of a detector. It is

\begin{eqnarray}
F_{pattern} (\hat{\mathbf{\Omega}}) = \mathbf{D} : \mathbf{e}_{pattern}(\hat{\mathbf{\Omega}})\, ,
\qquad \mathbf{D} =  \frac{1}{2}\left[ \hat{\mathbf{u}} \otimes \hat
{\mathbf{u}}- \hat{\mathbf{v}}
\otimes \hat{\mathbf{v}}\right]\:,
\label{eq2}
\end{eqnarray}
here we have $pattern=+,\times,s$.  The symbol ":"  is the contraction
between tensors.  $\mathbf{D}$  maps the  metric perturbation in a
signal on the detector. The vectors $\hat{\mathbf{u}}$ and
$\hat{\mathbf{v}}$ are unitary  and orthogonal to each other. They
are directed to each detector arm and form an orthonormal  system
with the unitary vector $\hat{\mathbf{w}}$.
$\hat{\mathbf{\Omega}}$ is the  unitary vector directed along the
GW propagation; $\hat{\mathbf{m}}$ and $\hat{\mathbf{n}}$ are
two orthonormal vectors (see Fig. \ref{fig2}). Eq.\,(\ref{eq2})
holds only when the arm length of the detector is smaller and
smaller than the GW wavelength that we are taking into account.
This is relevant to  deal with ground-based laser interferometers but this condition could not be valid
when dealing with space interferometers like LISA. However, more accurate studies have to be pursued on this issue.
A standard orthonormal coordinate system for the detector is
\begin{equation}
\left\{
\begin{array}{lll}
\hat{\mathbf{u}} = (1,0,0)
\\
\hat{\mathbf{v}} = (0,1,0)
\\
\hat{\mathbf{w}} = (0,0,1)\,.
\end{array}
\right. \;
\end{equation}
On the other hand,  the coordinate  system for the GW, rotated by
angles $(\theta, \phi)$,  is given by
\begin{equation}
\left\{
\begin{array}{lll}
\displaystyle
\hat{\mathbf{u}}^{\prime} =(\cos \theta \cos \phi , \cos \theta \sin \phi , -\sin \theta)
\\
\displaystyle
\hat{\mathbf{v}}^{\prime} = (- \sin \phi , \cos \phi , 0)
\\
\displaystyle
\hat{\mathbf{w}}^{\prime} = (\sin \theta \cos \phi , \sin \theta \sin \phi , \cos \theta)
\end{array}
\right. \;.
\end{equation}
The
rotation with respect to the angle $\psi$, around the
GW-propagating axis, gives the most general choice for the coordinate system, that is
\begin{equation}
\left\{
\begin{array}{lll}
\displaystyle
\hat{\mathbf{m}} = \hat{\mathbf{u}}^{ \prime} \cos \psi + \hat{\mathbf{v}}^{\prime} \sin \psi\\
\displaystyle
\hat{\mathbf{n}} = - \hat{\mathbf{v}}^{ \prime} \sin \psi + \hat{\mathbf{u}} ^{\prime} \cos \psi
\\
\displaystyle
\hat{\mathbf{\Omega}} = \hat{\mathbf{w}}^{ \prime}
\end{array}
\right. \;.
\end{equation}
Coordinates $(\hat{\mathbf{u}},\hat{\mathbf{v}},\hat{\mathbf{w}})$
are related to the coordinates
$(\hat{\mathbf{m}},\hat{\mathbf{n}},\hat{\mathbf{\Omega}})$ by the
rotation angles ($\phi,\,\theta,\,\psi$)as indicated in
Fig.\ref{fig2}. From the vectors $\hat{\mathbf{m}}$,
$\hat{\mathbf{n}}$, and $\hat{\mathbf{\Omega}}$, the polarization
tensors are
\begin{eqnarray}
\mathbf{e}_{+} &=& \frac{1}{\sqrt{2}}\left(\hat{\mathbf{m}} \otimes \hat{\mathbf{m}} -\hat{\mathbf{n}} \otimes \hat{\mathbf{n}}\right) \;, \\
\mathbf{e}_{\times} &=& \frac{1}{\sqrt{2}}\left( \hat{\mathbf{m}} \otimes \hat{\mathbf{n}} +\hat{\mathbf{n}} \otimes \hat{\mathbf{m}}\right) \;,  \\
\mathbf{e}_{s} &=& \frac{1}{\sqrt{2}}\left( \hat{\mathbf{\Omega}} \otimes \hat{\mathbf{\Omega}}\right) \;.  \end{eqnarray}
Taking into account  Eqs.\,(\ref{eq2}), the angular patterns for
each polarization are
\begin{eqnarray}
F_{+}(\theta, \phi, \psi) &=&  \frac{1}{\sqrt{2}} (1+ \cos ^2 \theta ) \cos 2\phi \cos 2 \psi- \cos \theta \sin 2\phi \sin 2 \psi \;, \label{5_1}\\
F_{\times}(\theta, \phi, \psi) &=& - \frac{1}{\sqrt{2}} (1+ \cos ^2 \theta ) \cos 2\phi \sin 2 \psi -  \cos \theta \sin 2\phi \cos 2 \psi \;, \label{5_2}\\
F_{s}(\theta, \phi) &=& \frac{1}{\sqrt{2}} \sin^2 \theta \cos 2\phi \;.
\label{5_3}
\end{eqnarray}
The angular pattern functions for each  polarization are plotted
in Figs.\,\ref{fig:3} and \ref{fig:3a}.  The figures show the sensitivity (i.e. the "gain") of the interferometer in a given direction. A typical antenna pattern  presents some maxima of sensitivity (lobes); the strongest of these maxima (the main lobe) defines the direction in which the antenna is most sensitive \cite{maggiore}. These considerations indicate  that the signal related to the scalar component (\ref{5_3}) is, in principle, detectable. Our results are consistent, for example,
with those in \cite{abio,nishi,tobar}.  Another possibility  to investigate the presence of  GW-scalar components is  to
consider the stochastic background of GWs as we will do in the next section.

%%%%%%%%%%%%%%%%%%%%%%%%%%%%%%%%%%%%%%%
 \section{GW-scalar component in the stochastic background}
 \label{sette}
 %%%%%%%%%%%%%%%%%%%%%%%%%%%%%%%%%%%%%

The primordial physical  process
can give rise to the characteristic spectrum $\Omega_{\varphi}$ for
the early stochastic background of relic scalar GWs. The
production  processes have been analyzed, for example, in
\cite{Allen,Grishchuk} but only for the   tensorial
components  related to the standard GR. Actually the process can be improved
considering also the scalar-tensor component considered here.

Before starting with the analysis, it has to be stressed that the stochastic background of scalar GWs related to the scalar field $\varphi$ can be characterized
by a dimensionless spectrum (see the analogous definition for
tensorial waves in
\cite{maggiore,Allen,Grishchuk,AO})\begin{equation}
\Omega_{\varphi}(f)=\frac{1}{\rho_{c}}\frac{d\rho_{\varphi}}{d\ln
f},\label{eq: spettro}\end{equation} where
 \begin{equation}
\rho_{c}\equiv\frac{3H_{0}^{2}}{8\pi G}\label{eq: densita'
critica}\end{equation} is the today critical energy density of
the Universe, $H_0$ the today observed Hubble expansion rate, and
$d\rho_{\varphi}$ is the energy density of the scalar part of the
gravitational radiation contained in the frequency range $f$ to
$f+df$. We are considering now standard units.

The existence of a relic stochastic background of scalar GWs is a
consequence of general assumptions. Essentially it derives from
basic principles of Quantum Field Theory and GR.
The strong variations of gravitational field in the early Universe
amplifies the zero-point quantum fluctuations and produces relic
GWs. It is well known that the detection of relic GWs is the only
way to learn about the evolution of the very early Universe, up to
the bounds of the Planck epoch and the initial singularity
\cite{maggiore,Allen,Grishchuk}. It is very important to stress
the unavoidable and fundamental character of such a mechanism. It
directly derives from the  inflationary scenario
\cite{Watson,Guth}, which well fit the WMAP data that are in particular
good agreement with almost exponential inflation and spectral
index $\approx1$ \cite{peacock}.

A remarkable fact about the inflationary scenario is that it
contains a natural mechanism which gives rise to perturbations for
any field. It is important for our aims that such a mechanism
provides also a distinctive spectrum for relic scalar GWs. These
perturbations, in inflationary cosmology, arise from the most basic
quantum mechanical effect: the uncertainty principle. In this way,
the spectrum of relic GWs that we could detect today is nothing
else but the adiabatically-amplified zero-point fluctuations
\cite{Allen,Grishchuk}. The calculation for a simple inflationary
model can be performed for the scalar field component. Let us assume that the early
Universe is described by an inflationary de Sitter phase emerging in
a radiation dominated phase \cite{Allen,Grishchuk,AO}. The
conformal metric element is
\begin{equation}
ds^{2}=a^{2}(\eta)[-d\eta^{2}+d\overrightarrow{x}^{2}+h^{(tot)}_{ij}(\eta,\overrightarrow{x})dx^{i}dx^{j}],\label{eq:
metrica}\end{equation} where, for a purely scalar GW the metric
perturbation (\ref{eq: perturbazionetotale}) reduces to
\begin{equation}
h^{(tot)}_{ij}=h_{\varphi} e_{ij}^{(s)}.\label{eq: perturbazione
scalare}\end{equation} Let us assume a phase transition between a
de Sitter  and a radiation-dominated phase \cite{Allen,Grishchuk},
we have: $\eta_1$ is the inflation-radiation transition conformal
time and $\eta_0$ is the value of conformal time today.  If we
express the scale factor in terms of comoving time
$cdt=a(t)d\eta$,
 we have
\begin{equation}
\label{eq:dominioradiazione}
a(t)\propto\exp(H_{ds}t), \qquad a(t)\propto\sqrt{t}
\end{equation}
 for the de Sitter and radiation phases respectively. In order to solve
 the horizon and flatness problems, the condition
${\displaystyle \frac{a(\eta_{0})}{a(\eta_{1})}>10^{27}}$ has to
be satisfied. The relic scalar-tensor GWs are the weak
perturbations $h^{(tot)}_{ij}(\eta,\overrightarrow{x})$ of the
metric (\ref{eq: perturbazione scalare}) which can be written in
the form

\begin{equation}
h^{(tot)}_{ij}=e_{ij}^{(s)}(\hat{k})X(\eta)\exp(i\overrightarrow{k}\cdot\overrightarrow{x}),\label{eq:
relic gravity-waves}\end{equation} in terms of the conformal time
$\eta$ where $\overrightarrow{k}$ is a constant wave-vector. From
Eqs. (\ref{eq: solS}) and (\ref{eq: relic gravity-waves}), the scalar component is
\begin{equation}
h_{\varphi}(\eta,\overrightarrow{k},\overrightarrow{x})=X(\eta)\exp(i\overrightarrow{k}\cdot\overrightarrow{x}),\label{eq:
phi}\end{equation} where we have specified the amplitude and the phase to the cosmological case.
Assuming $Y(\eta)=a(\eta)X(\eta)$, from the
Klein-Gordon equation in the FRW metric, one gets
\begin{equation}
Y''+(|k|^{2}-\frac{a''}{a})Y=0\label{eq:
Klein-Gordon}\end{equation} where the prime denotes the derivative
with respect to the conformal time. The solutions of Eq. (\ref{eq:
Klein-Gordon})  can be expressed in terms of   Hankel functions in
both the
inflationary and radiation dominated eras, that is:\\
for $\eta<\eta_{1}$ \begin{equation}
X(\eta)=\frac{a(\eta_{1})}{a(\eta)}[1+H_{ds}\omega^{-1}]\exp(-ik(\eta-\eta_{1})),\label{eq:
ampiezza inflaz.}\end{equation} for $\eta>\eta_{1}$
\begin{equation}
X(\eta)=\frac{a(\eta_{1})}{a(\eta)}\left[\alpha\exp(-ik(\eta-\eta_{1}))+\beta\exp
ik(\eta-\eta_{1})\right],\label{eq: ampiezza rad.}\end{equation}
where $\omega=ck/a$ is the angular frequency of the wave (which is
a function of  time only  being $k=|\overrightarrow{k}|$ constant);
$\alpha$ and $\beta$ are time-independent constants which we can
obtain demanding that both $X$ and $dX/d\eta$ are continuous at
the boundary $\eta=\eta_{1}$ between the inflationary and the
radiation dominated eras. By this constraint, we obtain
\begin{equation}
\alpha=1+i\frac{\sqrt{H_{ds}H_{0}}}{\omega}-\frac{H_{ds}H_{0}}{2\omega^{2}}\,,\qquad
\beta=\frac{H_{ds}H_{0}}{2\omega^{2}}\label{eq:
beta}\end{equation} In Eqs. (\ref{eq: beta}),
$\omega=ck/a(\eta_{0})$ is the angular frequency as observed
today, $H_{0}=c/\eta_{0}$ is the Hubble expansion rate as observed
today. Such calculations  are referred in  literature as the
Bogoliubov coefficient methods \cite{Allen,Grishchuk}.

In an inflationary scenario, every  classical or macroscopic
perturbation is damped out by the  inflation, i.e. the minimum
allowed level of fluctuations is that required by the uncertainty
principle. Solution (\ref{eq: ampiezza inflaz.}) corresponds
to a de Sitter vacuum state. If the period of inflation is long
enough, the today observable properties of the Universe  should be
indistinguishable from the properties of a Universe started in the
de Sitter vacuum state. In the radiation dominated phase, the
eigenmodes which describe particles are the coefficients of
$\alpha$ while  $\beta$- coefficients  describe antiparticles  (see also \cite{tuning}). Thus, the number
of  particles created at angular frequency $\omega$ in the
radiation dominated phase is
\begin{equation}
N_{\omega}=|\beta_{\omega}|^{2}=\left(\frac{H_{ds}H_{0}}{2\omega^{2}}\right)^{2}.\label{eq:
numero quanti}\end{equation}  Furthermore, it is possible to write an
expression for the energy density of the stochastic background  of  scalar relic gravitons in the frequency interval
$(\omega,\omega+d\omega)$ as
\begin{equation}
d\rho_{\varphi}=2\hbar\omega\left(\frac{\omega^{2}d\omega}{2\pi^{2}c^{3}}\right)N_{\omega}=
\frac{\hbar
H_{ds}^{2}H_{0}^{2}}{4\pi^{2}c^{3}}\frac{d\omega}{\omega}=\frac{\hbar
H_{ds}^{2}H_{0}^{2}}{4\pi^{2}c^{3}} \frac{df}{f}\,,\label{eq: de
energia}\end{equation} where $f$, as above, is the frequency in
standard comoving time. Eq. (\ref{eq: de energia}) can be
rewritten in terms of the today and de Sitter value of energy
density being
\begin{equation} H_{0}=\frac{8\pi G\rho_{c}}{3c^{2}}\,,\qquad H_{ds}=\frac{8\pi G\rho_{ds}}{3c^{2}}.\end{equation}
Introducing the Planck density ${\displaystyle
\rho_{Planck}=\frac{c^{7}}{\hbar G^{2}}}$, the spectrum is given by
\begin{equation}
\Omega_{\varphi}(f)=\frac{1}{\rho_{c}}\frac{d\rho_{\varphi}}{d\ln
f}=\frac{f}{\rho_{c}}\frac{d\rho_{\varphi}}{df}=\frac{16}{9}\frac{\rho_{ds}}{\rho_{Planck}}.\label{eq:
spettro gravitoni}\end{equation} At this point,  some  comments
are in order. First of all, such a calculation works for a
simplified model that does not include the matter dominated era.
If  such an era is also included, the redshift at equivalence
epoch has to be considered. Taking into account  results in
\cite{Allen}, we get
\begin{equation}
\Omega_{\varphi}(f)=\frac{16}{9}\frac{\rho_{ds}}{\rho_{Planck}}(1+z_{eq})^{-1},\label{eq:
spettro gravitoni redshiftato}\end{equation} for the waves which,
at the epoch in which the Universe becomes matter dominated, have
a frequency higher than $H_{eq}$, the Hubble parameter at
equivalence. This situation corresponds to frequencies
$f>(1+z_{eq})^{1/2}H_{0}$. The redshift correction in Eq.(\ref{eq:
spettro gravitoni redshiftato}) is needed since the today observed
Hubble parameter $H_{0}$ would result different  without a matter
dominated contribution. At lower frequencies, the spectrum is
given by \cite{Allen,Grishchuk}
\begin{equation}
\Omega_{\varphi}(f)\propto f^{-2}.\label{eq: spettro basse
frequenze}\end{equation} As a further consideration, let us note
that the results in Eqs. (\ref{eq: spettro gravitoni}) and (\ref{eq:
spettro gravitoni redshiftato}), which are not frequency
dependent, do not work correctly in all the range of physical
frequencies. For waves with frequencies less than today observed
$H_{0}$, the notion of energy density has no sense since the
wavelength becomes longer than the Hubble scale of the Universe.
In analogous way, at high frequencies, there is a maximal
frequency above which the spectrum rapidly drops to zero. In the
above calculation, the simple assumption that the phase transition
from the inflationary to the radiation dominated epoch is
instantaneous has been made. In the physical Universe, this
process occurs over some time scale $\Delta\tau$, being
\begin{equation}
f_{max}=\frac{a(t_{1})}{a(t_{0})}\frac{1}{\Delta\tau},\label{eq:
freq. max}\end{equation} which is the redshifted rate of the
transition. In any case, $\Omega_{\varphi}$ drops rapidly. The two
cutoffs at low and high frequencies for the spectrum guarantee
that the total energy density of the relic scalar gravitons is
finite. For Grand Unified Theories  energy-scale inflation, it is of the order
\cite{Allen}
\begin{equation}
\frac{\rho_{ds}}{\rho_{Planck}}\approx10^{-12}.\label{eq: rapporto
densita' primordiali}\end{equation} These results can be
quantitatively constrained considering the  WMAP release. In
fact, it is well known that WMAP observations put  severe
restrictions on the spectrum. Considering the ratio
$\rho_{ds}/\rho_{Planck}$, the relic scalar GW spectrum seems
consistent with the WMAP constraints on scalar perturbations.
Nevertheless, since the spectrum falls off $\propto f^{-2}$ at low
frequencies, this means that today, at LIGO-VIRGO and LISA
frequencies, one gets
\begin{equation} \Omega_{\varphi}(f)h_{100}^{2}<2.3\times
10^{-12}.\label{eq: limite spettro WMAP}\end{equation} It is
interesting to calculate the  corresponding strain at $\approx
100Hz$, where interferometers like VIRGO and LIGO reach a maximum
in sensitivity. The well known equation for the characteristic
amplitude \cite{Allen,Grishchuk} adapted to the scalar
component of GWs can be used:
\begin{equation}
h_{\varphi}(f)\simeq1.26\times
10^{-18}(\frac{1Hz}{f})\sqrt{h_{100}^{2}\Omega_{\varphi}(f)},\label{eq:
legame ampiezza-spettro}\end{equation} and then we obtain
\begin{equation}
h_{\varphi}(100Hz)<2\times 10^{-26}.\label{eq: limite per lo
strain}\end{equation}

Then, since we expect a sensitivity of the order of $10^{-22}$ for
the above interferometers at $\approx100Hz$, we need to gain four
order of magnitude. Let us analyze the situation also at smaller
frequencies. The sensitivity of the VIRGO interferometer is of the
order of $10^{-21}$ at $\approx10Hz$ and in that case it is
\begin{equation}
h_{\varphi}(100Hz)<2\times 10^{-25}.\label{eq: limite per lo
strain2}\end{equation} The sensitivity of the LISA interferometer
will be of the order of $10^{-22}$ at $\approx 10^{-3} Hz$ and in
that case it is
\begin{equation}
h_{\varphi}(100Hz)<2\times 10^{-21}.\label{eq: limite per lo
strain3}\end{equation} This means that a stochastic background of
relic scalar GWs could be, in principle, detected by the LISA
interferometer.
%%%%%%%%%%%%%%%%%%%%%%%%%%%%%%%%%%%%%%%%%%%%%%
\section{Discussion and conclusions}
\label{otto}
%%%%%%%%%%%%%%%%%%%%%%%%%%%%%%%%%
In this paper, we have investigated the possibility that
metric-affine $f(R)$ theories could be distinguished from purely
metric ones, by considering the difference in affine connections
which result in the relative variation of a suitable scalar field.
The absence of such a variation, that is ${\displaystyle
\frac{\delta \varphi}{\varphi}=0}$, is the indication that the
connection is Levi-Civita. In other words,  geodesic structure and causal
structure of the theory coincide.  On the other hand, if
${\displaystyle \frac{\delta \varphi}{\varphi}\neq0}$ a more
general theory, formulated onto a ${\bf U}_4$ manifold, has to be
considered. In principle, such an issue could be tested at a
fundamental level by detecting possible scalar modes of GWs. This is the key point of this paper.

In this perspective,
we have investigated the
detectability of additional polarization modes of a stochastic GW background
by ground-based laser-interferometric detectors and
space-interferometers. Such polarization modes, in general, appear
in the extended theories of gravitation and can be utilized to
constrain the theories beyond GR in a model-independent way.
However, a point has to be discussed in detail.  If the
interferometer is directionally sensitive and we also know the
orientation of the source (and of course if the source is
coherent) the situation is straightforward. In this case, the
massive mode coming from  $f(R)$-gravity
would induce longitudinal displacements along the direction of
propagation which should be detectable and only the amplitude due
to the scalar mode would be the true, detectable, "new" signal
\cite{greci}.
On the other hand, in the case of the stochastic background, there
is no coherent source and no directional detection of the
gravitational radiation. What the interferometer picks is just an
averaged signal coming from the contributions of all possible
modes from (uncorrelated) sources all over the celestial sphere.
Since we expect the background to be isotropic, the signal will be
the same regardless of the orientation of the interferometer, no
matter how or on which plane it is rotated, it would always record
the characteristic amplitude $h_\varphi$. So there is intrinsically no
way to disentangle any of the mode in the background, being $h_\varphi$
related to the total energy density of the gravitational
radiation, which  depends on the number of modes available.

This is the why we have considered only $h_{\varphi}$ in the above
cross-correlation analysis without giving further fine details
coming from polarization. For the situation considered here, we
have found that the massive modes are certainly of interest for
direct attempts of detection by the LISA experiment. It is, in
principle, possible that massive GW modes could be produced in
more significant quantities in cosmological or early astrophysical
processes in alternative theories of gravity. This situation
should be kept in mind when looking for a signature distinguishing
these theories from GR, and seems to deserve further
investigation. From our point of view, it is of extremely
importance since it could allow to  distinguish the same structure
of the space-time.

However,   some further considerations are   necessary at this point.
According to \cite{CNO},  $f(R)$-gravity in metric  formalism can be presented as an effective dark fluid assuming both the role of dark energy and dark matter.
Evidently, metric-affine $f(R)$-gravity may be also presented as (different) effective dark fluid. The above results could contribute in discriminating between the two approaches, if the signature of  affine connections is experimentally selected. In particular, torsion could have a relevant role in structure formation and in coincidence problem between dark energy and dark matter as discussed in \cite{CCSV1,CCSV2}. In this sense, dynamics is richer than that presented in \cite{CNO}. 
Furthermore,  it is well known that not all metric $f(R)$-gravity model may pass local tests, like Brans-Dicke test, Newton law, etc . The aim is to find out self-consistent models matching Solar System scales with extragalactic and cosmological scales.  In \cite{NO1,NO2}, criteria and several explicit models which pass local tests are given. In this paper, we have not faced this problem since we are using general $f(R)$-gravity models with the only request that they are analytic and Taylor expandable.  These  are sufficient conditions to discuss scalar GWs.  In order to construct  a reliable theory, also the problem of background has to be considered together with the stochastic GWs. This will be the argument of forthcoming researches aimed to achieve comprehensive and self-consistent models.

\end{document}